\documentclass[12pt]{article}
\usepackage{latexsym}
\usepackage{epsfig}

\usepackage{graphicx}
\usepackage{epstopdf}

\usepackage{epsfig,amssymb,amsmath,amscd}

\newcommand{\bq}{\begin{equation}}
\newcommand{\eq}{\end{equation}}
\newcommand{\bqa}{\begin{eqnarray}}
\newcommand{\eqa}{\end{eqnarray}}
\newcommand{\ben}{\begin{enumerate}}
\newcommand{\een}{\end{enumerate}}
\newcommand{\bc}{\begin{center}}
\newcommand{\ec}{\end{center}}
\newcommand{\bqb}{\begin{eqnarray*}}
\newcommand{\eqb}{\end{eqnarray*}}

\def\pr#1#2#3{Phys. Rev. ${\bf{#1}}$, #2 (#3)}

\def\pl#1#2#3{Phys. Lett. ${\bf{#1}}$, #2 (#3)}

\def\np#1#2#3{Nucl. Phys. ${\bf{#1}}$, #2 (#3)}

\def\jhep#1#2#3{JHEP ${\bf{#1}}$, #2 (#3)}
\def\epj#1#2#3{Eur. Phys. J. ${\bf{#1}}$, #2 (#3)}

\def\jmp#1#2#3{J. Mod. Phys. ${\bf{#1}}$, #2 (#3)}

\begin{document}
\pagenumbering{arabic}
\thispagestyle{empty}
\def\thefootnote{\fnsymbol{footnote}}
\setcounter{footnote}{1}

\begin{flushright}
Nov. 7, 2017\\
 \end{flushright}

\begin{center}
{\Large {\bf Tests of the Higgs and top quark properties through inclusive
$t\bar t$ production}}.\\
 \vspace{1cm}
{\large F.M. Renard}\\
\vspace{0.2cm}
Laboratoire Univers et Particules de Montpellier,
UMR 5299\\
Universit\'{e} Montpellier II, Place Eug\`{e}ne Bataillon CC072\\
 F-34095 Montpellier Cedex 5, France.\\
\end{center}

\vspace*{1.cm}
\begin{center}
{\bf Abstract}
\end{center}

We show how the inclusive $t\bar t$ production in $e^+e^-$, gluon~gluon
and photon~photon collisions could reveal the presence of new (visible or invisible)
particles connected to the Higgs sector and/or to the top quark.

\vspace{0.5cm}

\def\thefootnote{\arabic{footnote}}
\setcounter{footnote}{0}
\clearpage

\section{INTRODUCTION}

Possible compositeness of the top quark and of the Higgs boson
is an interesting question which motivated many works, see  
\cite{partialcomp,Hcomp2,Hcomp3,Hcomp4}.
The observability of top compositeness has also been discussed in
\cite{Tait}.\\
We have recently analyzed possible tests of the concept of
Compositeness Standard Model (CSM), see ref.\cite{CSMrev},
using several Higgs boson and top quark production processes.\\
Detailed analyses of processes involving Higgs boson and/or top quark
could reveal their specific properties but first indications about such
a composite nature could simply appear in inclusive distributions
through departures from SM predictions. We have given examples 
with single top, $H$ or $Z$ inclusive distributions.\\

In this paper we extend this type of study to inclusive
$t\bar t$ production. The $M_X$ distribution of the $t\bar t +X$
production opens directly the view to new states, with any kind
of quantum numbers, even invisible ones, to which it may be a portal.\\

In SM at lowest order, $X$ can be a Higgs boson, a photon or a $Z$ boson;
see figs.1,2,3.\\

In this first exploration we ignore electroweak and QCD radiative corrections;
we just want to explore what gross new features would be generated
by Higgs compositeness.\\

New states may first consist in "excited" states originating
from a substructure (see for example \cite{comp})
or from the SM extension with a new sector.
They can involve a new $H'$ or a new $Z'$. In SM the $G^0$ is equivalent 
to the longitudinal $Z_L$. With the CSM concept this may also be the case
for the excited states with a $G'^0$ equivalent to a $Z'_L$.\\

A more complex possibility is multibody $X$ production.
In SM it would be suppressed by higher order $\alpha$ or 
$\alpha_s$ factors. This suppression would not occur if the multibody
production originates from substructure recomposition (like
in hadronic production after creation of basic quarks and
antiquarks) or from the connection to a new strongly 
interacting sector to which the Higgs boson and/or the
top quark may be a portal.\\

We will give examples of such effects with arbitrary values of masses
and couplings.\\

In the following sections we give illustrations 
of the inclusive $d\sigma/dM_X$ distribution
at a fixed total energy in $e^+e^-$, gluon~gluon and photon~photon collisions.\\
Apart from (not shown) standard narrow  peaks at $M_X=m_H,m_Z$  one should 
observe broad peaks at $m_{H'}$,  $m_{Z'}$ (or $m_{G^{'0}}$) for "excited"
states and thresholds for (visible or invisible) multiparticle production.
We will conclude this note by mentioning that its main purpose 
is to motivate further phenomenological and experimental studies of the
inclusive $t\bar t$ production.\\

\section{Analyses of $e^+e^-,\gamma\gamma,gg\to t\bar t  X$}

\subsection{$e^+e^- \to t\bar t  X$}

As explained in the introduction the basic SM terms of the $M_X$ distribution
correspond to Higgs, photon and $Z$ boson emission as shown in Fig.1.\\ 
In the case of the Higgs boson the emission occurs from the $t$ and $\bar t$
final lines or from $ZH$ production followed by $Z\to t\bar t$.\\
Photon and  $Z$  production occurs similarly to the $H$ case but with, in addition,
the emission from $e^+$ and $e^-$ lines. Photon and $Z$ have transverse components 
and the $Z$ can also have longitudinal components ($Z_L$) equivalent,
at high energy, to Goldstone $G^0$ production through diagrams
exactly similar to those of the Higgs case.\\ 

Effect of Higgs and top compositeness modifying the involved couplings has been
discused in previous studies, see \cite{CSMrev},
and illustrated for example through the
introduction of effective form factors keeping the SM structure at low energies,
like
\bq
F(s)={s_0+M^2\over s+M^2}~~\label{FF}
\eq
where $s_0$ is a threshold and $M$ a new physics scale taken for example equal
to $1$ TeV.
See also \cite{trcomp} where the corresponding concept of effective mass has 
been introduced. 
We will use it for showing the change produced in the inclusive
distribution when such a form factor affects each Higgs and Goldstone coupling,
replacing $s$ by $M^2_X$ in the above equation.\\

But several types of new terms may also appear.\\ 

A first type may consist of the occurence of "excited" states.
The case of an "excited" Higgs boson $H'$ is illustrated 
with an arbitrary mass $M_{H'}=700$ GeV, a width $\Gamma_{H'}=0.1$ GeV and couplings 
to Z and top quark similar to those of the basic Higgs boson.\\
In the same spirit one may expect the occurence of an "excited" $Z'$, with $Z'_L$
equivalent to an "excited" Goldstone state $G'^0$.
We illustrate its presence with a mass $M_{G^{'0}}=M_{Z'}=1$ TeV, 
a width $\Gamma_{G^{'0}}=\Gamma_{Z'}=0.1$ GeV and couplings to Z and top similar to those of the
standard $G^0$. The diagrams would be the same as those of Fig.1 for $H$ and $Z$ production.
replacing $H$ by $H'$ (similar diagrams for $G'^0$) and $Z$ by $Z'$.\\

In the next step we assume the existence of a new  (visible or invisible) sector
coupled to the Higgs or directly to the top quark.
It may be a strongly coupled sector or the result of a substructure which
creates multibody production similarly to the case of hadronic production generated
after quark+antiquark creation.
As an example we will assume that, after their production
according to the diagrams of Fig.1, the Higgs or the Goldstone boson
creates a pair of subconstituents with a mass $m_0=1$ TeV. Multiparticle production 
then occurs automatically with a threshold at $M_X=2m_0=2$ TeV.\\

We have computed the total $M_X$ distribution of the inclusive $t\bar t$ production
$d\sigma/dM_X$ by integrating the differential cross section over the energies 
and angles of the $t$ and the $\bar t$ with $M_X^2=(q-p_t-p_{\bar t})^2$ where $q$
is the total center of mass momentum, $s=q^2$.\\

Arbitray cuts have been used in order to avoid collinear singlularities.  The
absolute values of the cross sections have no predictive meaning, we only want to
discuss shapes which may be typical of the various above dynamical assumptions.

These shapes of the $M_X$ distribution are shown in Fig.4 for a total center 
of mass energy of $\sqrt{s}=5$ TeV.\\

The basic SM shape is only due to Higgs, photon and $Z$ boson emission
(the trivial peaks at  $m_{H}$ and $m_{Z}$ are not shown).
New effects appear clearly with 
bumps at $m_{H'}$ , $m_{G^0}$ and a multiparticle threshold at $2m_0$.\\

We also show how these shapes would
be modified by the presence of an effective form factor $F(M^2_X)$ affecting the 
Higgs and Goldstone couplings.\\

To appreciate these effects in an other way we illustrate the corresponding
ratios of the new contributions (SM+$H$ sector, SM+$G$ sector, SM+$H+G$ sectors)
over the pure SM one.\\

\subsection{$gluon~gluon \to t\bar t  X$}

The lowest order SM diagrams are shown in Fig.2. The Higgs boson, photon and $Z$ 
can only be emitted by the final $t$ and  $\bar t$ lines.  Contrarily to the $e^+e^-$ 
case there is no emission from the initial lines at this order.\\
At this step of our study we ignore the QCD corrections and in particular the gluon emission.\\
As above we then add the contributions of excited 
$H'$ and $G^{'0}$ states and the continuum of multiparticle
production from the $H$ and $G^0$ sectors.\\
In both cases we also consider the effects of a form factor affecting the
$H$ and $G^0$ couplings as well as those of the excited states.\\ 
The corresponding illustrations are given in Fig.5 for $\sqrt{s}=15$ TeV.
Bumps and threshold effects are qualitatively similar to those of 
the $e^+e^-$ case.\\

\subsection{$\gamma\gamma \to t\bar t  X$}

The corresponding SM diagrams are shown in in Fig.3.  The Higgs boson, photon and $Z$ 
can again only be emitted by the final $t$ and  $\bar t$ lines. 
The global features are rather similar to those of the $gluon~gluon \to t\bar t  X$
process.\\
The addition of the excited states and multibody production from
the new Higgs and Goldstone sectors generates bumps and threshold effects
as shown in Fig.6.\\
Form factor effects are alo similar to those of the previous cases.\\

\section{Conclusion}

We have shown that the inclusive $t\bar t$ production in 
$e^+e^-$, gluon~gluon and photon~photon  collisions may give
remarkable signals of new physics properties of the Higgs boson
and/or the top quark.\\
We have illustrated the possibility of "excited" $H'$, $Z'$ (or $G^{'0}$) states
and of (parton like) multiparticle production above some threshold.
In each case specific departures with respect to SM expectation
could be observed.\\
These illustrations may motivate further phenomenological 
and experimental studies in this respect.
For what concerns the experimental domain, 
in $e^+e^-$ collisions see \cite{Moortgat, Denterria, Craig, Englert}, 
in hadronic collisions see \cite{Contino,Richard} and in 
photon-photon collisions see \cite{gammagamma}.\\

Such an inclusive study should be especially fruitful if the new sectors connected to
Higgs or Goldstone consist, at least partly, of unvisible particles.\\
Confirmations of a possible signal could then be searched
in other Higgs and Goldstone production processes.
Exclusive processes for example $e^+e^-\to G'^0H'$ or $e^+e^-\to G'^+G'^-$
generalizing the standard $e^+e^-\to ZH$ or $e^+e^-\to W^+W^-$
could be interesting but difficult to analyse if invisible particles
are involved.\\
From the phenomenological side if such a signal would appear,
specific models could be tested and precise expectations should
be computed for well defined multibody
final states taking into account higher order SM coreections of 
electroweak or QCD nature.\\

\newpage

\begin{figure}[p]
\[
\epsfig{file=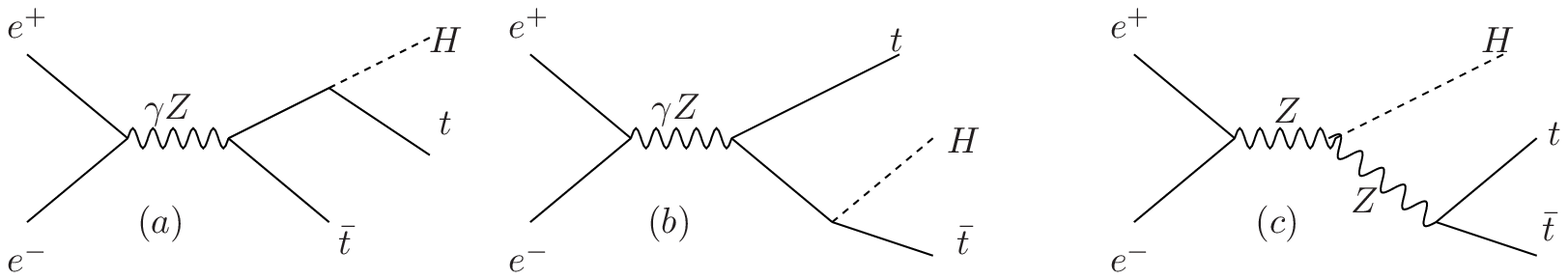 , height=3.cm}
\]\\
\[
\epsfig{file=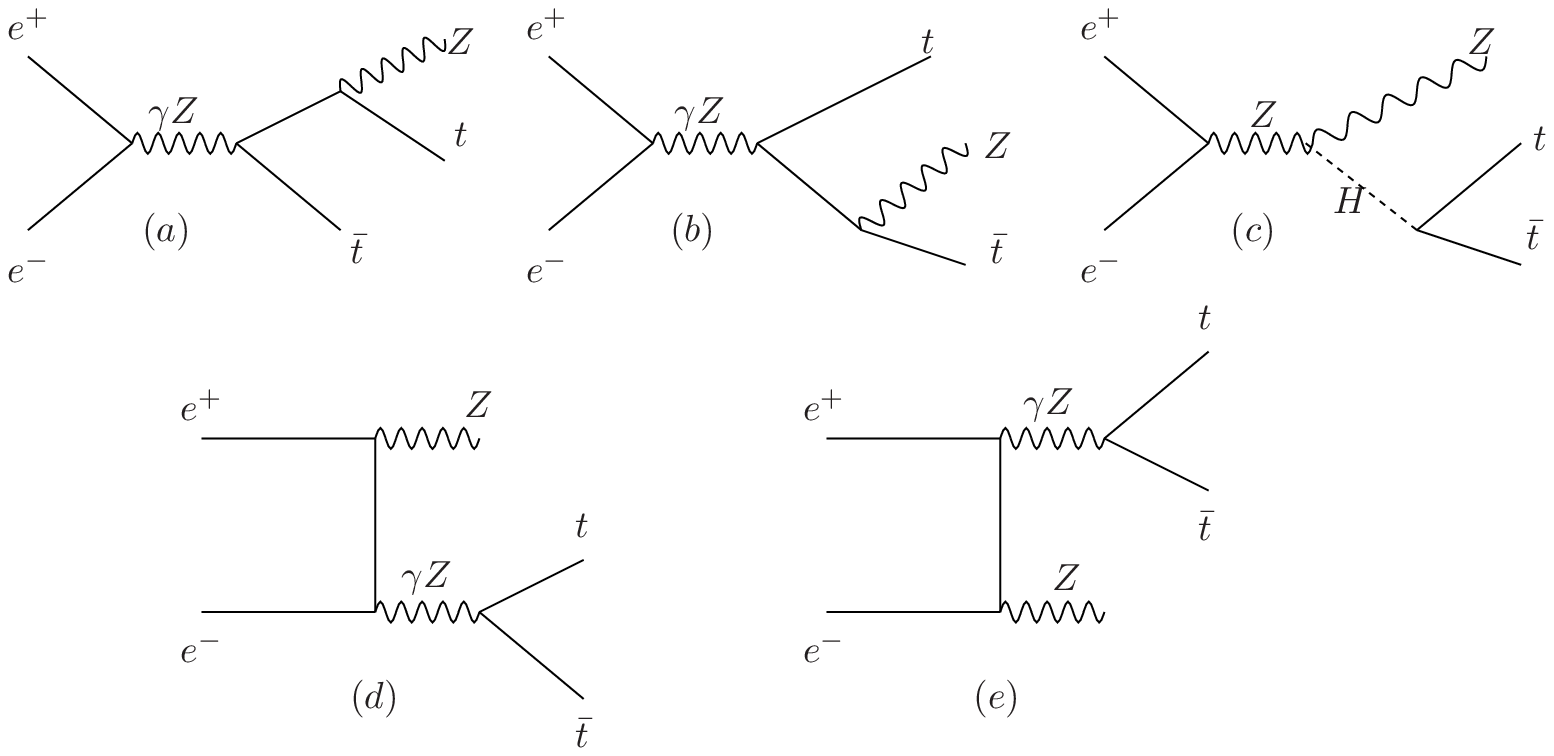, height=7.cm}
\]\\
\caption[1] {Lowest order SM diagrams for $e^+e^- \to t\bar t  H$
and $e^+e^- \to t\bar t  Z$ (and similarly for $e^+e^- \to t\bar t  \gamma$) .}
\end{figure}

\begin{figure}[p]
\[
\epsfig{file=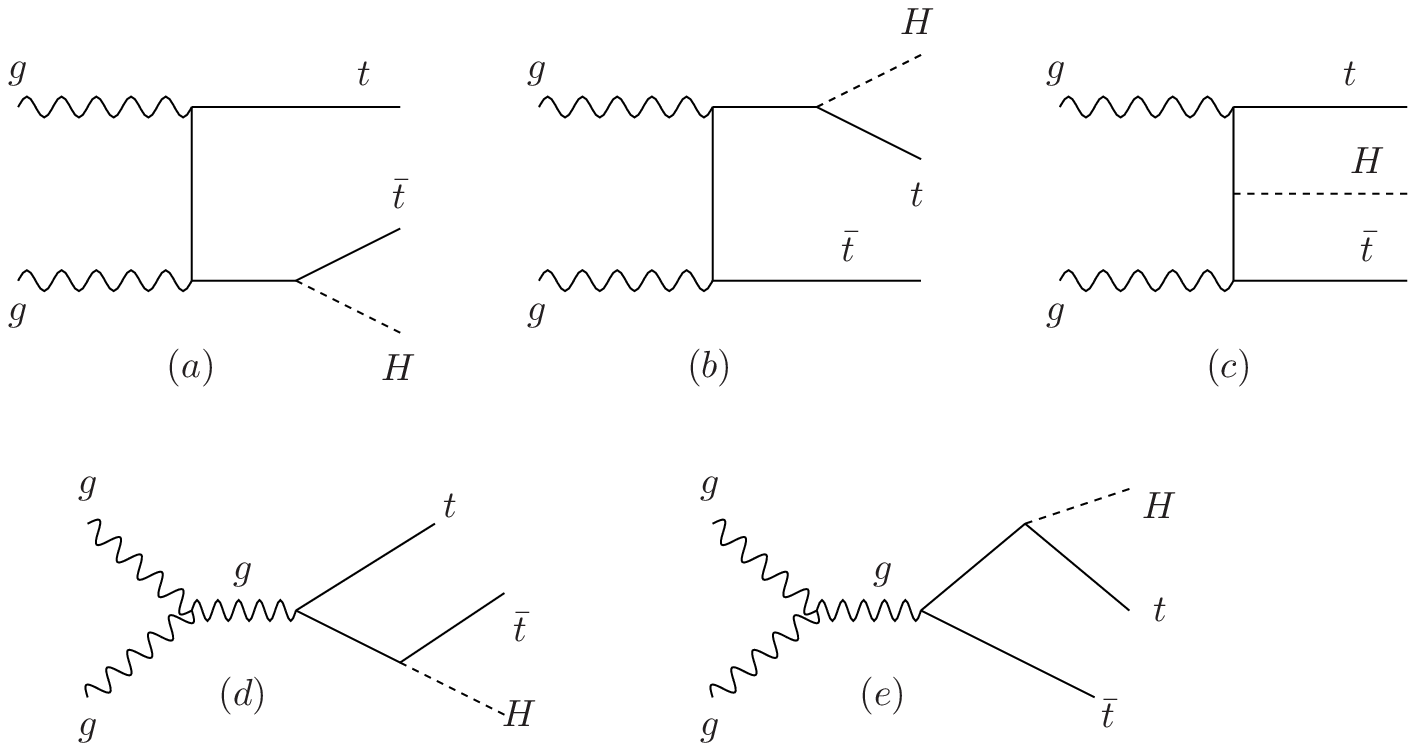 , height=8.cm}
\]\\
\[
\epsfig{file=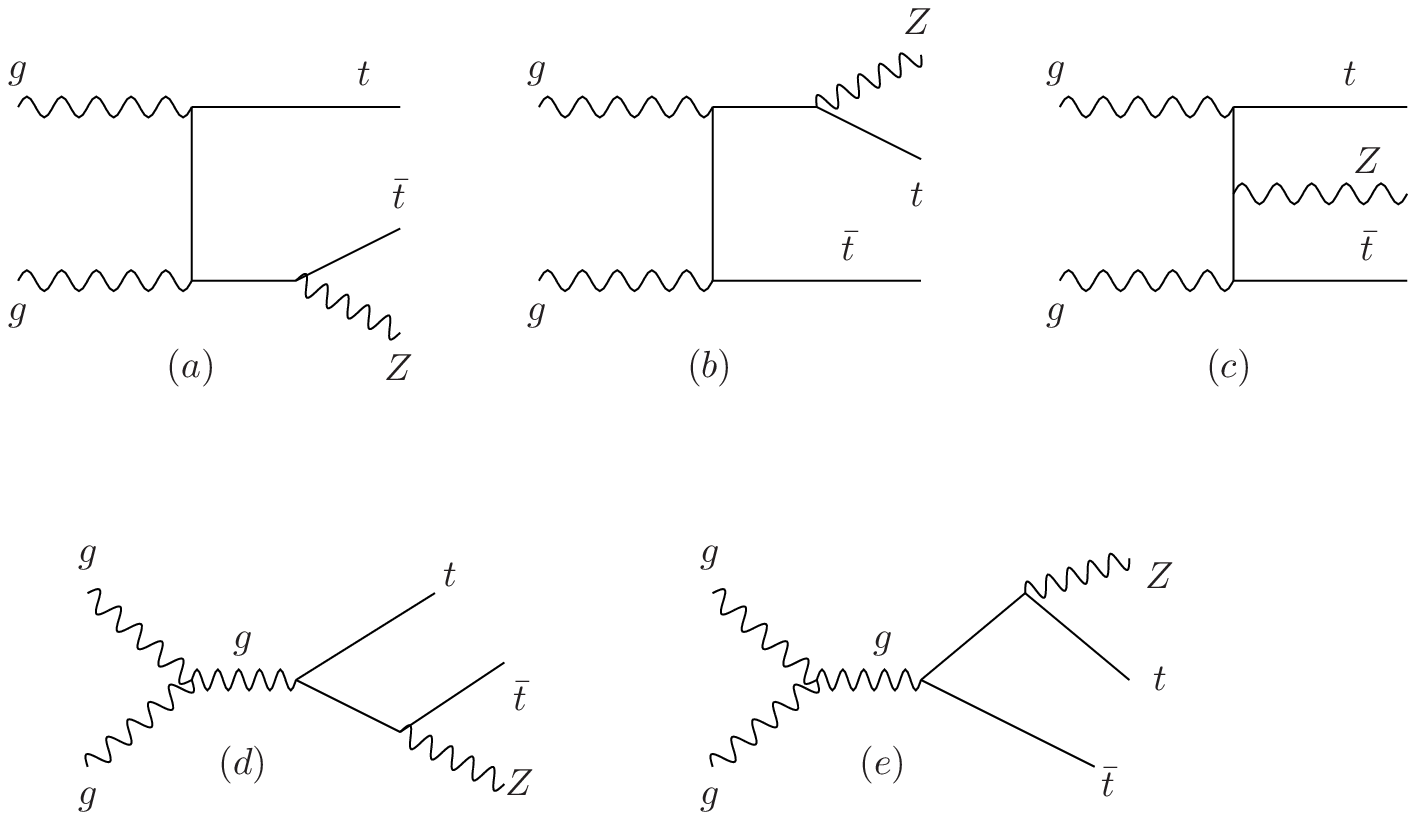, height=7.cm}
\]\\
\caption[1] {Lowest order SM diagrams for $gluon~gluon\to t\bar t  H$
and $gluon~gluon \to t\bar t  Z$ (and similarly for $gluon~gluon \to t\bar t  \gamma$).}
\end{figure}

\begin{figure}[p]
\[
\epsfig{file=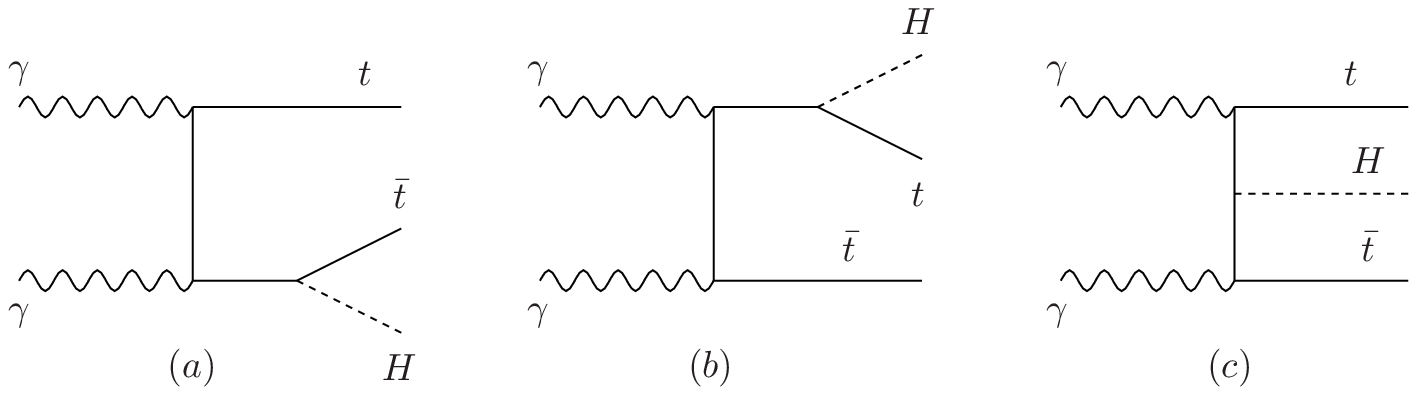 , height=4.cm}
\]\\
\[
\epsfig{file=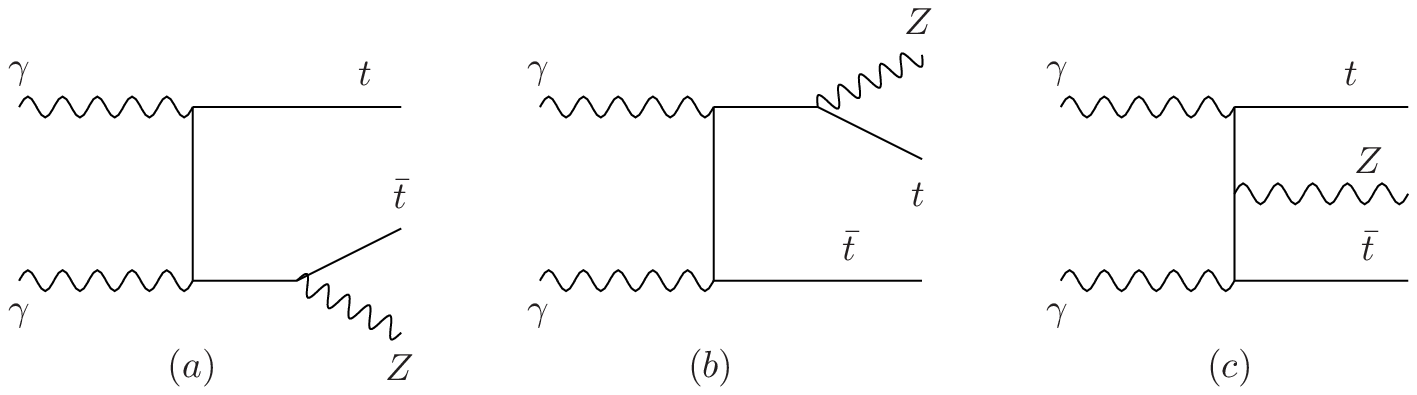, height=4.cm}
\]\\
\caption[1] {Lowest order SM diagrams for $\gamma\gamma\to t\bar t  H$
and $\gamma\gamma \to t\bar t  Z$ (and similarly for $\gamma\gamma \to t\bar t  \gamma$).}
\end{figure}

\begin{figure}[p]
\[
\epsfig{file=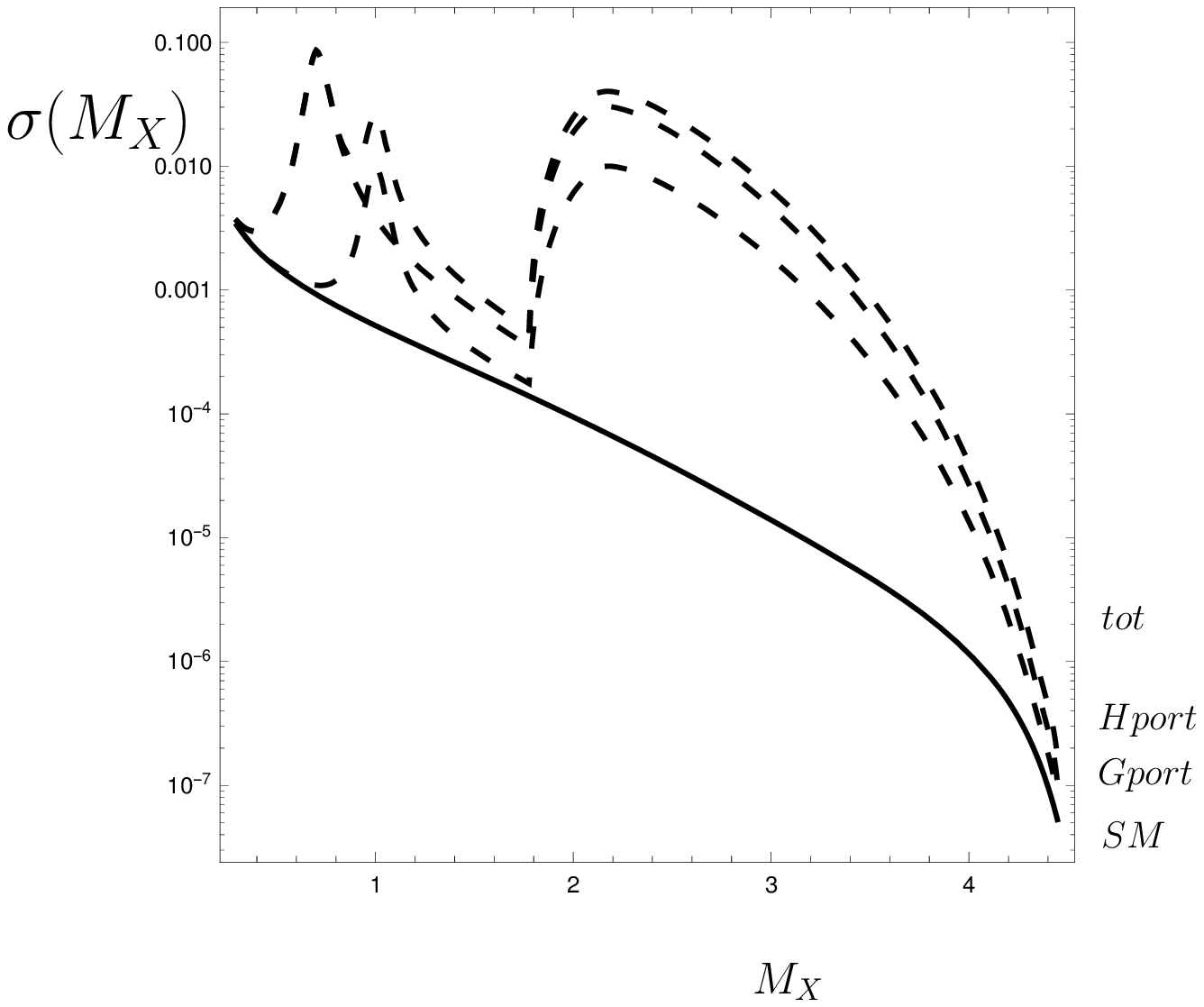, height=7.cm}
\hspace{0cm}
\epsfig{file=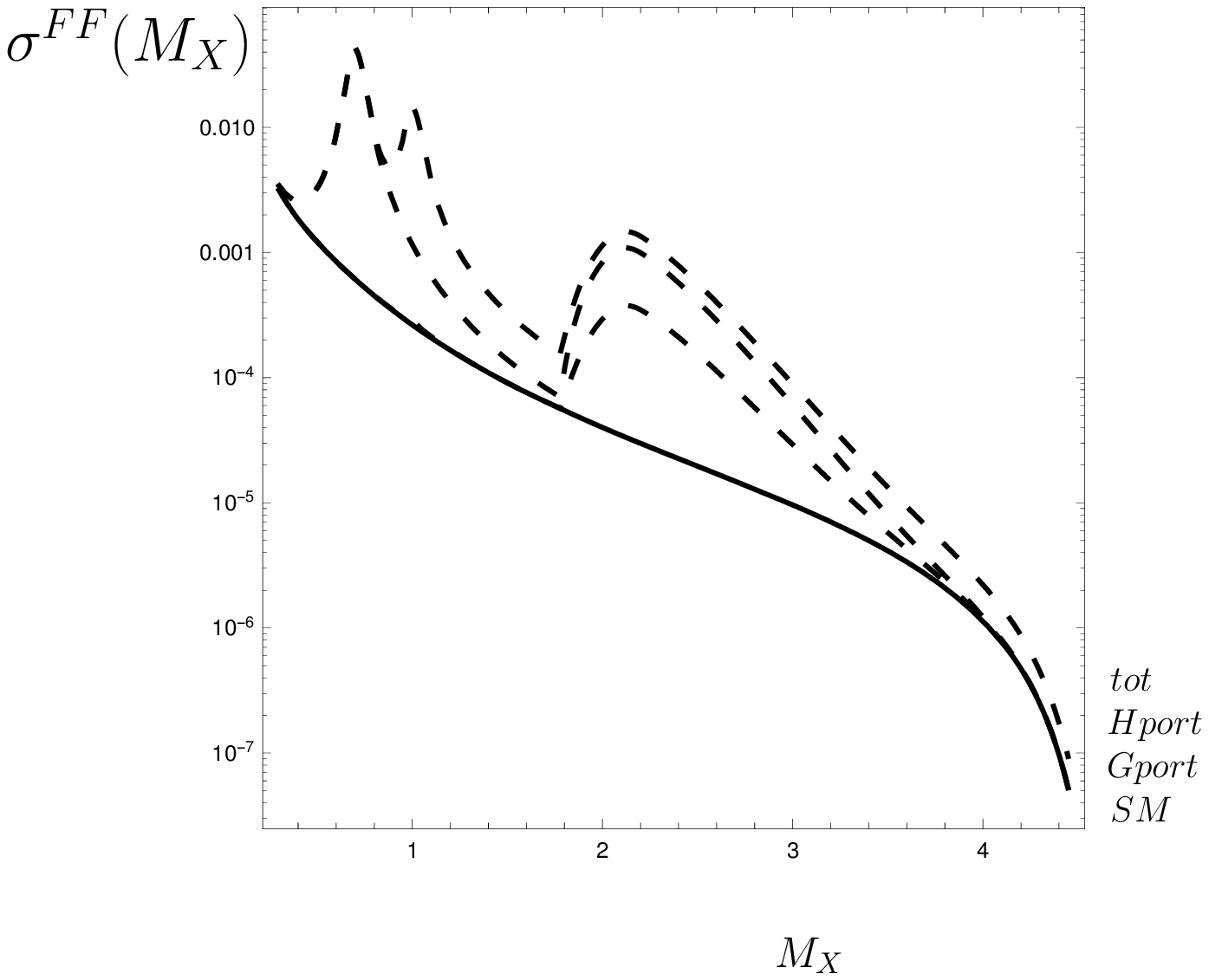, height=7.cm}
\]\\
\[
\epsfig{file=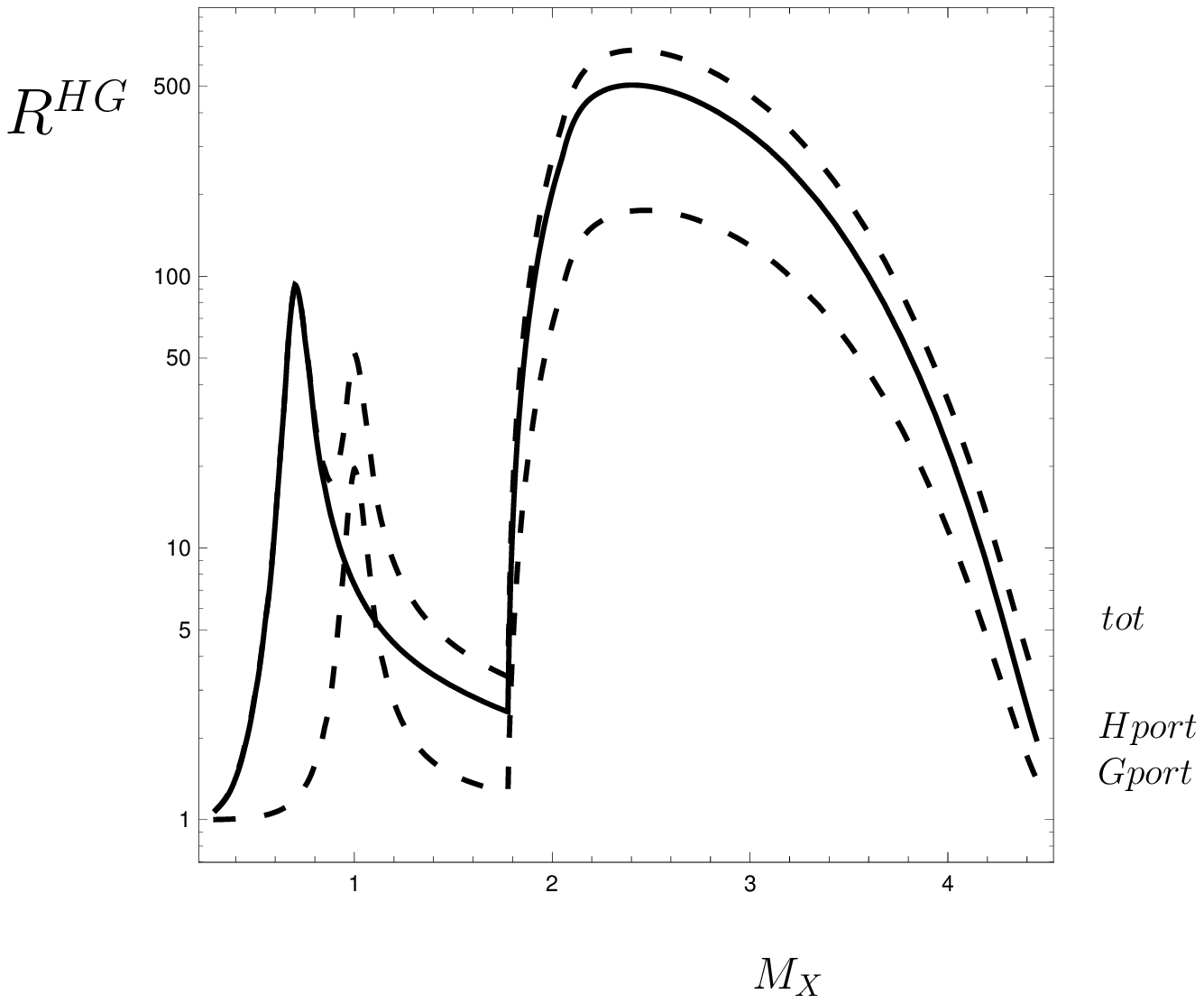, height=7.cm}
\hspace{0cm}
\epsfig{file=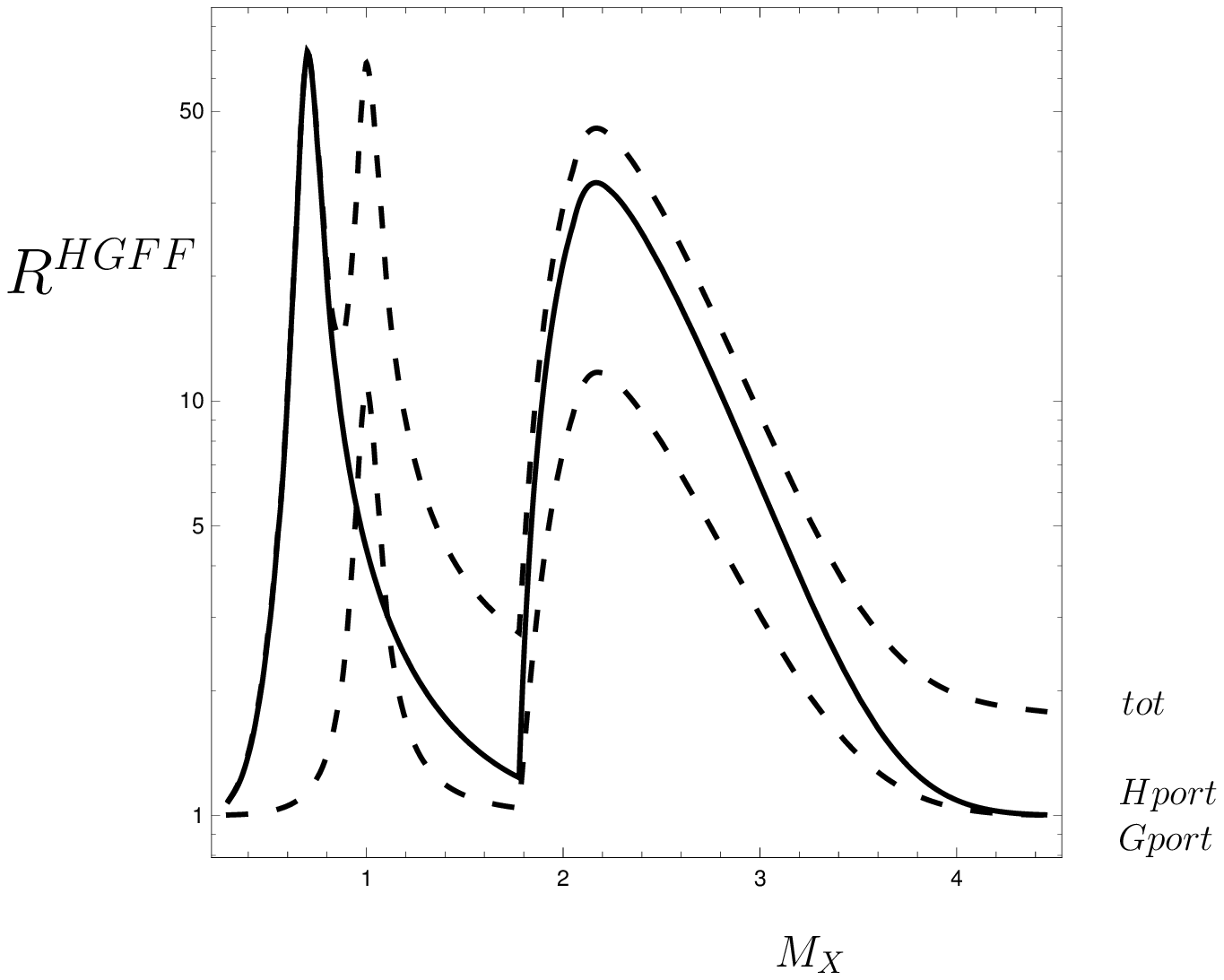, height=7.cm}
\]\\
\caption[1] {$M_X$ distribution of the $e^+e^-\to t\bar t  X$ cross section
(upper level) without(left) and with(right) form factor
and of the ratios over SM case of $H$ or $G$ new sector effects (lower level)
without(left) and with(right) form factor.}
\end{figure}

\begin{figure}[p]
\[
\epsfig{file=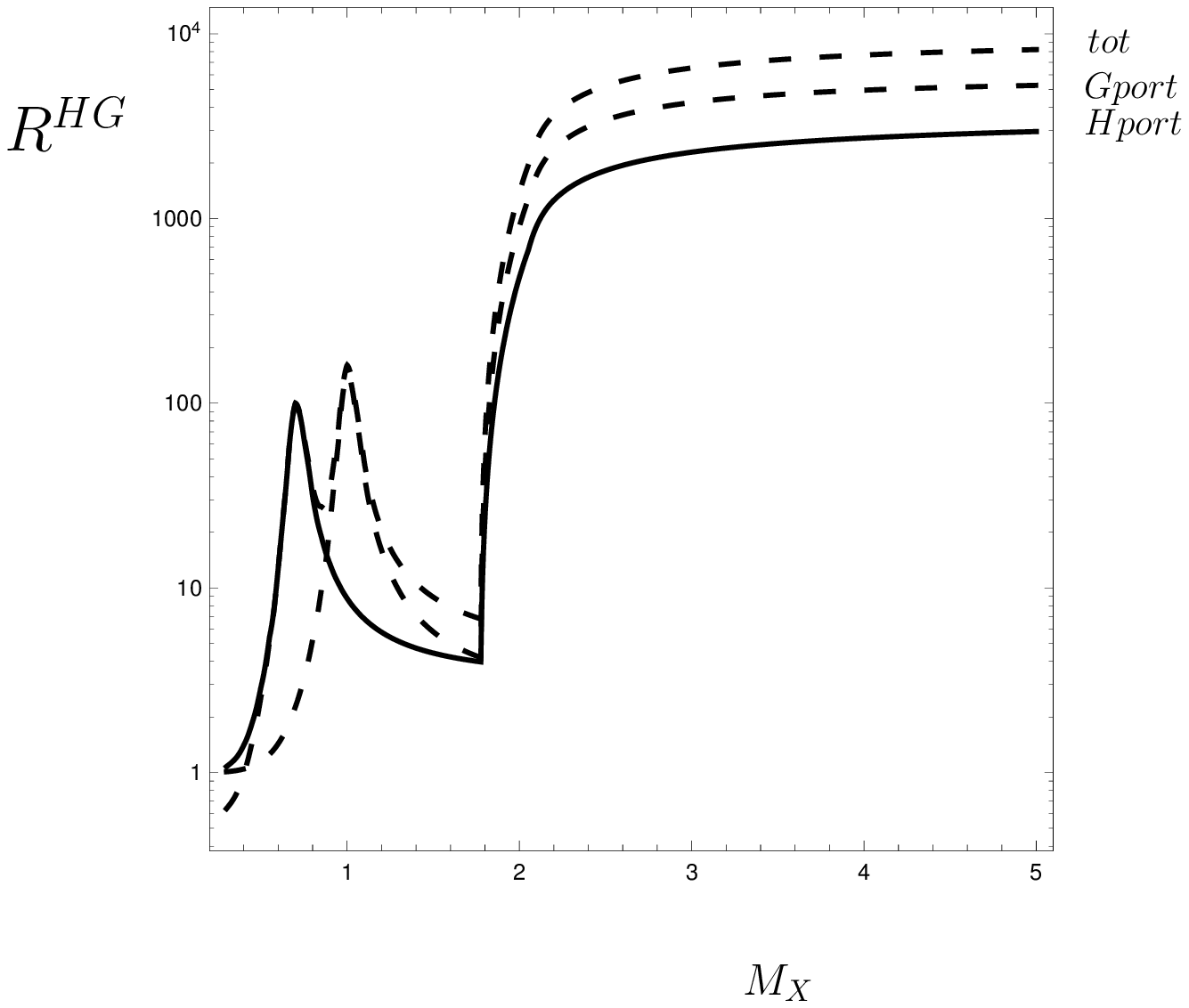, height=8.cm}
\]\\
\[
\epsfig{file=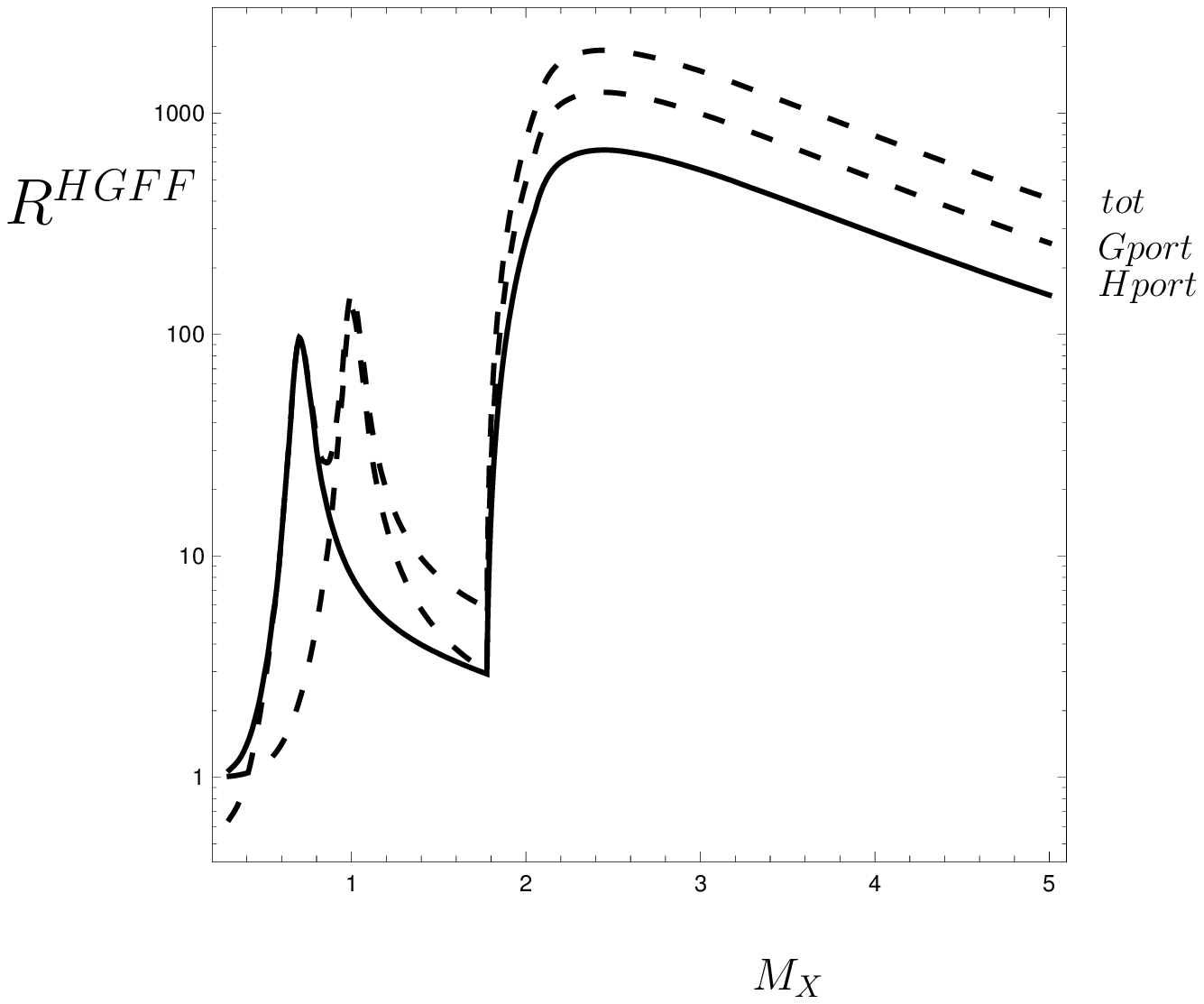, height=8.cm}
\]\\
\caption[1]  {$M_X$ distribution of the $gluon~gluon\to t\bar t  X$ cross section;
ratios over SM case of $H$ or $G$ new sector effects
without(upper level) and with(lower level) form factor.}
\end{figure}
\clearpage

\begin{figure}[p]
\[
\epsfig{file=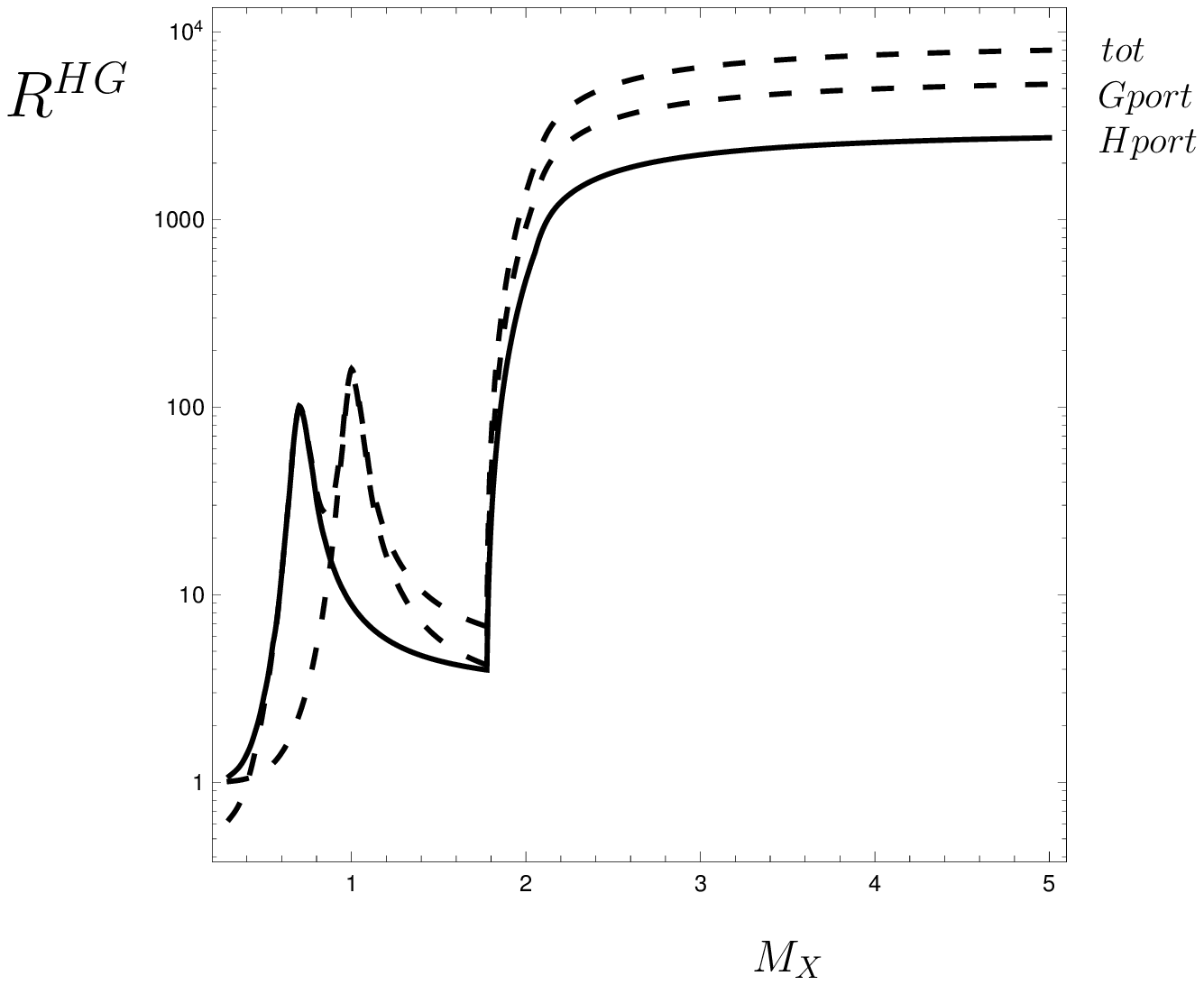, height=8.cm}
\]\\
\[
\epsfig{file=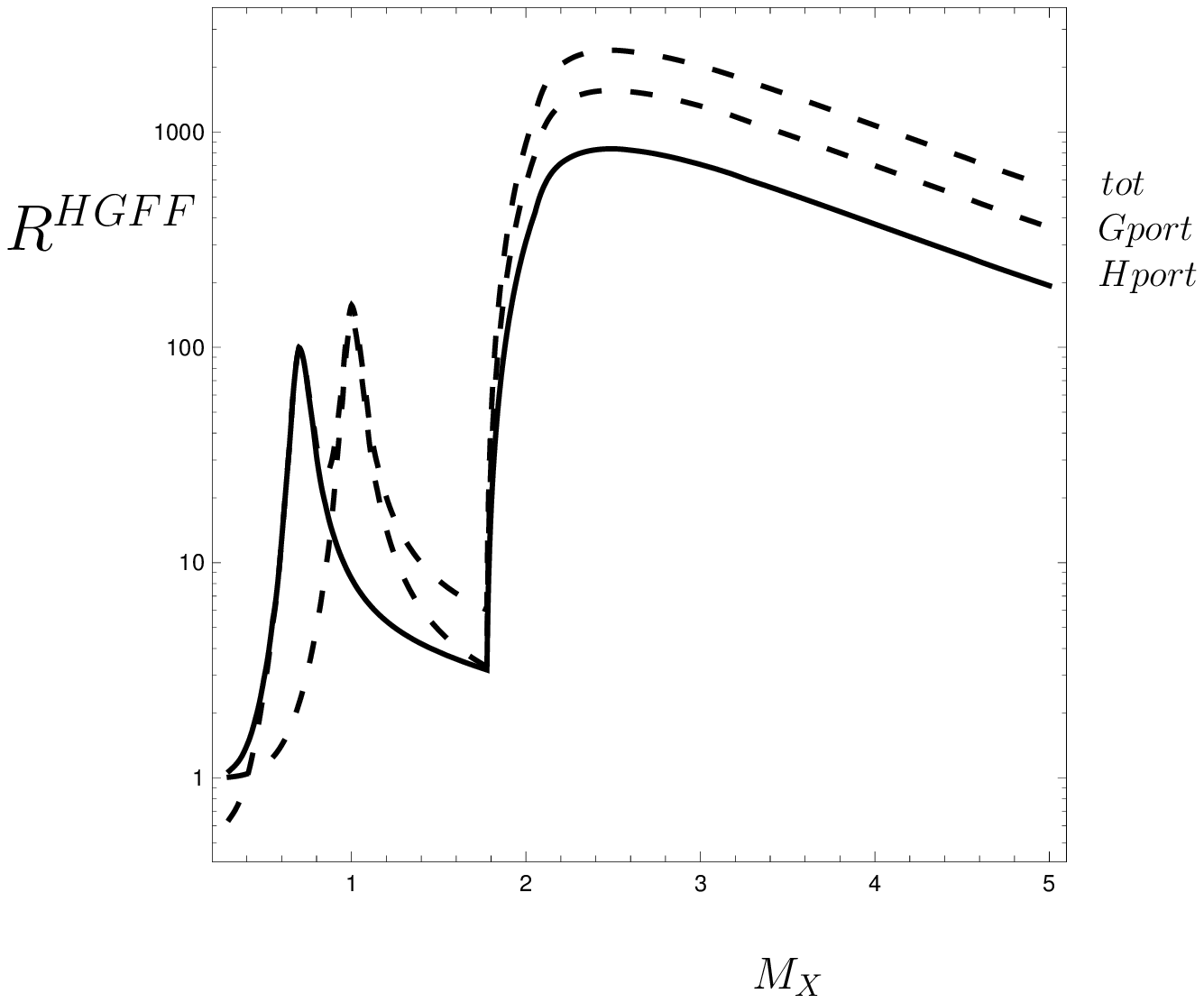, height=8.cm}
\]\\
\caption[1]  {$M_X$ distribution of the $\gamma\gamma\to t\bar t  X$ cross section;
ratios over SM case of $H$ or $G$ new sector effects
without(upper level) and with(lower level) form factor.}
\end{figure}

\end{document}